\documentstyle[12pt]{article}

\def\ar{\rightarrow}
\def\bib{\bibitem}

\def\intX{\int\! d^{D}X\,}

\def\lar{\longrightarrow}

\def\lp{l_{_P}}
\def\mp{m_{_P}}
\def\pa{\partial}
\def\rvec{\!\!\!\!^{^\rightarrow}}

\def\Tr{\,\mbox{Tr}\,}

\def\ga{\gamma}
\def\de{\delta}
\def\ep{\varepsilon}

\def\la{\lambda}

\def\si{\sigma}

\def\Ga{{\it\Gamma}}
\def\La{{\it\Lambda}}
\def\Om{{\it\Omega}}

\def\PSI{{\it\Psi}}

\def\Th{{\it\Theta}}

\def\beq{\begin{equation}}
\def\eeq{\end{equation}}
\def\bed{\begin{displaymath}}
\def\eed{\end{displaymath}}
\def\beqq{\begin{eqnarray}}
\def\eeqq{\end{eqnarray}}
\def\bedd{\begin{eqnarray*}}
\def\eedd{\end{eqnarray*}}

\begin{document}

\centerline{\normalsize\bf ISOMETRODYNAMICS AND GRAVITY}

\vspace*{0.9cm}
\centerline{\footnotesize C. WIESENDANGER}
\baselineskip=12pt
\centerline{\footnotesize\it Aurorastr. 24, CH-8032 Zurich}
\centerline{\footnotesize E-mail: christian.wiesendanger@zuerimail.com}

\vspace*{0.9cm}
\baselineskip=13pt
\abstract{Isometrodynamics (ID), the gauge theory of the group of volume-preserving diffeomorphisms of an "inner" $D$-dimensional flat space, is tentatively interpreted as a fundamental theory of gravity. Dimensional analysis shows that the Planck length $\l_P$ - and through it $\hbar$ and $\Ga$ - enters the gauge field action linking ID and gravity in a natural way. Noting that the ID gauge field couples solely through derivatives acting on "inner" space variables all ID fields are Taylor-expanded in "inner" space. Integrating out the "inner" space variables yields an effective field theory for the coefficient fields with $\l_P^2$ emerging as the expansion parameter. For $\hbar \ar 0$ only the leading order field does not vanish. This classical field couples to the matter Noether currents and charges related to the translation invariance in "inner" space. A model coupling this leading order field to a matter point source is established and solved. Interpreting the matter Noether charge in terms of gravitational mass Newton's inverse square law is finally derived for a static gauge field source and a slowly moving test particle. Gravity emerges as potentially related to field variations over "inner" space and might microscopically be described by the ID gauge field or equivalently by an infinite string of coefficient fields only the leading term of which is related to the macroscopical effects of gravity.}

\vspace*{0.7cm}

{\footnotesize\noindent{\it (A scientist) appears as realist insofar as he seeks to describe a world independent of the acts of perception; as idealist insofar as he looks upon the concepts and theories as the free invention of the human spirit (not logically derivable from what is empirically given); as positivist insofar as he considers his concepts and theories justified only to the extent to which they furnish a logical representation of relations among sensory experiences. He may even appear as Platonist or Phytagorean insofar as he considers the viewpoint of logical simplicity as an indispensable and effective tool of his research\footnote{Albert Einstein, in G. Holton and Y. Elkana (Eds.), {\it Albert Einstein - Historical and Cultural Perspectives} (Princeton University Press, 1982).}.}

\normalsize\baselineskip=15pt

\section{Introduction}
One of the great unsolved problems in physics is the formulation of a consistent quantum theory of gravity.

In the search of candidate theories for the description of gravity at both the classical and quantum level and guided by the principles that spacetime is a four-dimensional Minkowski manifold, that all fundamental interactions in Nature are governed by gauge symmetries and that quantum field theories of the fundamental interactions are renormalizable we have developed Isometrodynamics in \cite{chw2, chw3}. This is the gauge theory of the group ${\overline{DIFF}}\,{\bf R}^D$ of volume-preserving or "isometric" diffeomorphisms of an "inner" flat space $({\bf R\/}^{D},g)$ endowed with an a priori metric $g$.

In \cite{chw2} we have developed Classical Isometrodynamics within both the Lagrangian and the Hamiltonian frameworks and demonstrated that the theory can be formulated with a rigour similar to that achieved for classical Yang-Mills gauge field theories \cite{stw2}.

In \cite{chw3} we have developed Quantum Isometrodynamics and shown that it can be formulated as a renormalizable, asymptotically free gauge field theory with a formal rigour again similar to the one achieved for Yang-Mills gauge theories of compact Lie groups \cite{stw2}.

By its very definition Isometrodynamics fulfills important requirements towards any theory of gravity such as universality, i.e. the universal coupling of gravity to all fundamental fields.

So can we develop Isometrodynamics into a serious candidate theory to describe gravity consistently at both the classical and quantum level? This is what we will discuss in the following.

The notations and conventions used are given in Appendix A.

\section{Isometrodynamics}
In this section we review the basics of Classical Isometrodynamics \cite{chw2}.

Let us start with some definitions and conventions. In this paper we will work with fields $\PSI (x,X)$ defined on the product of four-dimensional Minkowski spacetime $x^\mu \in ({\bf M\/}^{\sl 4},\eta)$ and an "inner" $D$-dimensional Euclidean space $X^M \in ({\bf R\/}^{D},g)$ endowed with a flat a priori metric $g_{MN}$. By definition $Riem(g)=0$ so there always exist global coordinates with $g_{MN} = \de_{MN}$. The fields might carry additional representations of compact Lie groups such as the color $SU(3)$ related to the strong interaction. Further notations and conventions are given in Appendix A.

Isometrodynamics is the gauge theory of the infinite-dimensional group ${\overline{DIFF}}\,{\bf R}^D$ of all volume-preserving or "isometric" diffeomorphisms acting on $({\bf R\/}^{D},g)$.

In analogy to non-Abelian gauge theories the local invariance under infinitesimal "isometric" diffeomorphisms of "inner" space
\beq \label{1}
X^N\lar X'^N = X^N + {\cal E}^N (x,X),\quad \nabla_N {\cal E}^N (x,X) = 0
\eeq
requires the introduction of a covariant derivative and a gauge field
\beq \label{2}
D_\mu (x,X)\equiv \pa_\mu + A_\mu (x,X),
\quad A_\mu (x,X)\equiv A_\mu\,^M (x,X)\cdot \nabla_M
\eeq
with components $A_\mu\,^M$ transforming inhomogenously under local gauge transformations
\beq \label{3}
\de_{_{\cal E}} A_\mu\,^M = \pa_\mu {\cal E}^M + A_\mu\,^N \cdot
\nabla_N {\cal E}^M - {\cal E}^N \cdot \nabla_N A_\mu\,^M
\eeq
to ensure the covariant transformation of $D_\mu$.

The related field strength operator is defined by
\beq \label{4}
F_{\mu\nu} (x,X)\equiv [D_\mu (x,X), D_\nu (x,X)]
\eeq
and its components w.r.t. $\nabla_M$
\beq \label{5}
F_{\mu\nu}\,^M = \pa_\mu A_\nu\,^M - \pa_\nu A_\mu\,^M
+ A_\mu\,^N \cdot \nabla_N A_\nu\,^M
- A_\nu\,^N \cdot \nabla_N A_\mu\,^M
\eeq
transform homogenously
\beq \label{6}
\de_{_{\cal E}} F_{\mu\nu}\,^M = F_{\mu\nu}\,^N \cdot \nabla_N {\cal E}^M - {\cal E}^N \cdot \nabla_N F_{\mu\nu}\,^M.
\eeq

It is crucial that $A_\mu$ and $F_{\mu\nu}$ can be decomposed w.r.t $\nabla_M$. They are elements of the gauge algebra ${\overline{\bf diff}}\,{\bf R}^D$ and their components $A_\mu\,^M$ and $F_{\mu\nu}\,^M$ as well as ${\cal E}^M$ fullfil
\beq \label{7}
\nabla_M f^M = 0
\eeq
valid for all algebra elements $f^M$ ensuring volume preservation. 

There are only two relevant representations of the gauge algebra. First we have the scalar representation with covariant derivative
\beq \label{8}
D_\mu = \pa_\mu + A_\mu\,^M\cdot \nabla_M
\eeq
in which all fields live with the exception of the gauge fields introduced above. In the sequel we will call them "matter" fields. Second, there is the vector representation with covariant derivative
\beq \label{9}
{\cal D}_\mu^M\,_N = \pa_\mu \, \de^M\,_N + A_\mu\,^L \cdot 
\nabla_L \,\de^M\,_N - \nabla_N A_\mu\,^M
\eeq
in which the gauge fields and field strength components live. Note that the "inner" metric has played no role so far in the definition of the theory.

The Lagrangian for Isometrodynamics in natural units is given by \cite{chw2}
\beq \label{10}
L_{ID} (A_\nu\,^M, \pa_\mu A_\nu\,^M, \nabla_N A_\nu\,^M, \La)
\equiv \frac{1}{4}\,\frac{D(D+2)}{\Om_D}\,\Tr_\La \Big\{F_{\mu\nu}\, F^{\mu\nu}\Big\}_g,
\eeq
where $\Om_D\equiv\frac{2\pi^{D/2}}{(2\pi)^D\Ga(D/2)}$ and the trace operation $\Tr_\La\{\dots\}_g$ involving the "inner" metric $g_{MN}$ and a new parameter $\La$ has been properly introduced in \cite{chw2}. The parameter $\La$ carries dimension of momentum and necessarily occurs in Isometrodynamics to define a dimensionless volume element for "inner" space integrals. As shown in \cite{chw2} we can reduce the trace to
\beq \label{11}
L_{ID} (A_\nu\,^M, \pa_\mu A_\nu\,^M, \nabla_N A_\nu\,^M, \La)
= -\frac{\La^2}{4} \intX \La^D\, F_{\mu\nu}\,^M \cdot F^{\mu\nu}\,_M.
\eeq
"Inner" indices in general gauges are contracted with $g_{MN}$.

Note that rescaling $\La\ar \rho\La$ the resulting Lagrangian is related to the original one by \cite{chw2}
\beq \label{12}
L_{ID} (X,A_\nu\,^M(X),\dots, \rho\La) =
L_{ID} (\rho X,\rho A_\nu\,^M(X),\dots, \La).
\eeq
In other words theories for different $\La$ are equivalent up to "inner" rescalings.

The classical dynamics following from the Lagrangian Eqn.(\ref{10}) which governs the evolution of the gauge fields in four-dimensional spacetime has been discussed in detail in \cite{chw2} and shown there to be independent of the "inner" metric $g_{MN}$.

To couple any type of "matter" field the minimal coupling prescription suggests (1) to allow the "matter" fields to live on ${\bf M\/}^{\sl 4}\times {\bf R\/}^{D}$ - adding the necessary additional "inner" degrees of freedom -  and (2) to replace ordinary derivatives through covariant ones $\pa_\mu\ar D_\mu$ in the "matter" Lagrangians. This prescription leads to a universal coupling of any such "matter" field to the gauge fields of Isometrodynamics.

In \cite{chw2,chw3} we have Isometrodynamics shown to be a consistent theory at both the classical and the quantum level - more specifically pure Quantum Isometrodynamics (QID) is perturbatively renormalizable and asymptotically free, which we have explicitly demonstrated at the one-loop level. Coupled to the Standard Model (SM) fields QID is asymptotically free only for $D \geq 7$ - it is not for $2 \leq D \leq 6$ and we expect observable asymptotic states of QID to exist for those dimensions of "inner" space.

\section{Viable Theories of Gravity}
In this section we review the basic requirements towards viable theories of gravity as formulated in \cite{cmw}.

Let us start with the Dicke framework stating fundamental assumptions behind the formulation of any theory in physics \cite{cmw}. These are that (A) spacetime is a four-dimensional differentiable manifold whose points correspond to physical events and that (B) physical equations can be expressed in a form that is independent of the particular coordinates used.

The first assumption obviously holds for Isometrodynamics. W.r.t. the second assumption some comments are in place. We have developed Isometrodynamics on the same footing as the other gauge theories of fundamental interactions, i.e. in a Poincar\'e-invariant way which is per definition not covariant under general spacetime coordinate transformations. In analogy to the coordinate-independent reformulation of other gauge theories we could restate Isometrodynamics in a generally covariant way. However, the dynamical content of the theory is completely contained in its gauge invariance formulated on flat spacetime. Rewriting it in general coordinates would not add any physical content, but rather obscure its essence.

In addition it is important to note that based on Wigner's theorem  all the relativistic QFT-notions such as "particle", "one-particle state", "one-particle wave function" etc are deeply interlinked with Poincar\'e invariance and can be consistently defined only within the realm of relativistic QFT on a flat background geometry \cite{stw1}. As a result of Isometrodynamics being formulated on Minkowski spacetime these notions can immediately be generalized to the context of QID.

Let us next cite four basic criteria for the viability of a theory of gravity as stated in \cite{cmw} against which we will check Isometrodynamics below. They are:

{\it (i) A theory of gravity must be complete, i.e. it must be capable of analysing from "first principles" the outcome of any experiment of interest

(ii) A theory of gravity must be self-consistent, i.e. predictions for the outcome of an experiment must not depend on the approach of calculating them if there are different ways

(iii) A theory of gravity must be relativistic, i.e. in the limit as gravity is "turned off" the nongravitational laws of physics must reduce to the laws of special relativity

(iv) A theory of gravity must have the correct Newtonian limit, i.e. in the limit of weak gravitational fields and slow motions, it must reproduce Newton's law.}

To discuss these criteria in the context of Isometrodynamics we note that - very much unlike in the usual approach which starts with the discussion of classical point particles interacting with gravity - the only way to formulate an interacting theory is by starting from a comprehensive field-theoretical description of Nature such as the Standard Model coupled to ID through the minimal coupling prescription given in Section 2. By definition this is a {\it complete, self-consistent and relativistic} framework describing the interaction of the Standard Model fields with the gauge fields of Isometrodynamics - which validates criteria (i) - (iii). Below we will further comment on the technical implementation of (iii). On top there are good reasons to assume that the combined SM-ID model is renormalizable \cite{chw3} beyond one loop.

As a consequence, the question of what interaction in Nature - if any - is described by Isometrodynamics boils down to whether we can interpret the theory in terms of gravitational physics and uncover a Newtonian limit to validate (iv).

\section{Isometrodynamics in SI units}
In this section we establish Isometrodynamics in SI units.

Starting with the fact that $({\bf R\/}^{D},g)$ is a metric space it is natural to measure $X$ in units of meters, $[X]$ = m and its Fourier-conjugate variable $K$ in units of momentum, $[K]$ = kg $\cdot$ m $\cdot$ s$^{-1}$ = N $\cdot$ s. Note that $\La$ has the same units $[\La]$ = N $\cdot$ s as $K$. As a consequence the product of $X$ and $K$ is measured in units of an action, $[K\cdot X]$ = N $\cdot$ m $\cdot$ s = J $\cdot$ s = $[\hbar]$. The dimension of $d^D X\, \La^D$ is $[d^D X\cdot \La^D] = [\hbar]^D$ and a dimensionless integration measure over ${\bf R\/}^{D}$ is given by
\beq \label{13}
d^D X\, \left(\frac{\La}{\hbar}\right)^D.
\eeq
This measure is unique up to a dimensionless factor which remains unspecified because of the possibility of global "inner" rescalings \cite{chw2}.

Next we have $[D_\mu] = [\pa_\mu]$ = m$^{-1}$ or $[A_\mu\,^K\cdot \nabla_K]$ = m$^{-1}$ which means that $A_\mu\,^K$ is dimensionless, $[A_\mu\,^K]$ = 1. As a consequence the field strength components are measured in $[F_{\mu\nu}\,^K]$ = m$^{-1}$.

The Lagrangian density ${\cal L}_{ID}$ of ID from Eqn.(\ref{11}) is proportional to
$\left(\frac{\La}{\hbar}\right)^2 \, F_{\mu\nu}\,^M \cdot F^{\mu\nu}\,_M$  and is measured in units of $[\frac{\La}{\hbar}]^2 \, [F_{\mu\nu}\,^M]^2$ = m$^{-4}$. As $[\hbar c]$ = N $\cdot$ m$^2$ we set
\beq \label{14}
{\cal L}_{ID} \propto -\, \hbar c\, \left(\frac{\La}{\hbar}\right)^2 \, F_{\mu\nu}\,^M \cdot F^{\mu\nu}\,_M
\eeq
up to a dimensionless factor to ensure that ${\cal L}_{ID}$ has the correct dimension $[{\cal L}_{ID}]$ = N $\cdot$ m$^{-2}$. The integration over "inner" space does not change this because the integration measure Eqn.(\ref{13}) is dimensionless.

Finally, $\La$ carries units of momentum and we can take it proportional to a mass times a velocity built from the fundamental constants of nature $\hbar$, $c$ and $\Ga$. So we set $\La$ equal to the Planck mass $\mp = \sqrt{\frac{\hbar c}{\Ga}} = 2.17\times 10^{-8}$ kg times the velocity of light $c$,
\beq \label{15}
\La \equiv \mp c.
\eeq
Note that
\beq \label{16}
\frac{\La}{\hbar} = \sqrt{\frac{c^3}{\hbar \Ga}} = \frac{1}{\lp}
\eeq
then turns out to be the inverse Planck length $\lp = 1.62\times 10^{-35}$ m.

As a consequence we have
$\hbar c\, \left(\frac{\La}{\hbar}\right)^2 = \frac{c^4 }{\Ga}$ and find
\beq \label{17}
{\cal L}_{ID} \propto -\, \frac{c^4}{\Ga} \: F_{\mu\nu}\,^M \cdot F^{\mu\nu}\,_M,
\eeq
up to a dimensionless factor and independent of $\hbar$. Taking this factor as $\frac{1}{4 g^2}$ where $g$ is the dimensionless coupling of Isometrodynamics we finally have
\beq \label{18}
{\cal L}_{ID} = -\,\frac{1}{4 g^2}\, \frac{c^4}{\Ga} \: F_{\mu\nu}\,^M \cdot F^{\mu\nu}\,_M.
\eeq

The Planck mass, length and time are the only entities carrying unit of mass, length and time which can be built from the three basic constants $\hbar$, $c$ and $\Ga$ in Nature. Hence, from the dimensional analysis above the gravitational constant enters in a natural and up to rescalings unique way into ID.

Note that $\hbar$ does not enter Eqn.(\ref{18}) as we would expect for a field theory which should have a classical limit. Finally, it is important to note that $g$ is the fundamental coupling constant of ID - and not $\Ga$ - and that $g$ can be arbitrarily rescaled at the classical level - another reminder of the scale invariance of ID Eqn.(\ref{12}).

As a result the action of Isometrodynamics in SI units up to an arbitrary "inner" rescaling reads
\beq \label{19}
S = - \int \!dt\int d^{\sl 3}x\int \!d^D X\, \lp^{-D}\, \frac{1}{4 g^2}\, \frac{c^4}{\Ga} \: F_{\mu\nu}\,^M \cdot F^{\mu\nu}\,_M
\eeq 
and is measured in the correct units $[S] = [t][x]^3[{\cal L}]$ = N $\cdot$ m $\cdot$ s of an action.

\section{Effective Field Theory for Field Modes of Isometrodynamics related to a Power Series Expansion in $X$-space}
In this section we analyze the field modes of Isometrodynamics related to an expansion of the gauge parameter, gauge fields and the field strength components into power series' in "inner" space. The expansion starts off with the translation and rotation modes related to the local Euclidean subgroup $E({\bf R}^D)$ of the gauge group ${\overline{DIFF}}\,{\bf R}^D$. Expanding the action of Isometrodynamics and integrating out the "inner" space variables we derive an effective field theory for the various field modes organized in even powers of the Planck length which singles out the dynamics related to the translation mode as the leading term.

To get a better grasp on the physics potentially described by Isometrodynamics let us Taylor-expand all the gauge parameter, the gauge fields and field strength components in $X$-space up to $O(X^2)$. The intuition behind this is that keeping the leading constant term in such a Taylor expansion only and neglecting any field variations (i.e. $X$-derivatives) around fields constant over "inner" space is tantamount to completely decoupling ID fields from any other field and to neglecting all ID field self-couplings because all interactions with the ID gauge field come in through field derivatives $A_\mu\,^M \cdot {\nabla\rvec}_M$. Only if we allow for field variations in "inner" space do we get both the ID selfinteraction as well as the interaction of the ID gauge fields with other fields. Below we will show that we can formalize this idea to recover the usual Standard Model from the Standard Model coupled to full Isometrodynamics as the leading term in such an effective field theory expansion and validate (iii) in turn.

As a matter of convention we will denote by lower case letters the $x$-dependent Taylor coefficients of all the gauge parameter, the gauge fields and field strength components of ID denoted in full by capital letters.

Let us start with the expansion of the gauge parameter
\beqq \label{20}
{\cal E}^M (x,X) &=& \ep^M (x) + \ep^M\,_R (x)\,X^R \nonumber \\
&+& \ep^M\,_{RS} (x)\,X^R X^S + O(X^3)
\eeqq
keeping in mind $\nabla_M {\cal E}^M = 0$. The first two terms are related to the local Euclidean subgroup $E({\bf R}^D)= T(D)\times O(D)$ of the gauge group ${\overline{DIFF}}\,{\bf R}^D$ with $\ep^M (x)$ denoting local translations and $\ep_{asym}^{MN} (x)$ local rotations, where
\beq \label{21}
\ep^{MN} (x) = \ep_{asym}^{MN} (x) + \ep_{sym}^{MN} (x),
\quad \ep_{sym}^M\,_M (x) = 0
\eeq
is decomposed in its anti-symmetric and symmetric components.

In an identical way we expand the gauge fields
\beqq \label{22}
A_\mu\,^M (x,X) &=& a_\mu\,^M (x) + a_\mu\,^M\,_R (x)\,X^R \nonumber \\
&+& a_\mu\,^M\,_{RS} (x)\,X^R X^S + O(X^3).
\eeqq
The first two terms are the translation and rotation field modes $a_\mu\,^M$ and $a_\mu\,_{asym}^{MN}$ respectively related to the invariance of Isometrodynamics under the local Euclidean group $E({\bf R}^D)$, where
\beq \label{23}
a_\mu\,^{MN} (x) = a_\mu\,_{asym}^{MN} (x) + a_\mu\,_{sym}^{MN} (x),
\quad a_\mu\,_{sym}^M\,_M (x) = 0
\eeq
is again decomposed in its anti-symmetric and symmetric components.

From Eqn.(\ref{3}) we next read off the transformation behaviour of the Taylor coefficients Eqn.(\ref{22}) under gauge transformations Eqn.(\ref{20})
\beqq \label{24}
\de_{_{\ep}} a_\mu\,^M &=& \pa_\mu \ep^M + \ep^M\,_N \cdot a_\mu\,^N - a_\mu\,^M\,_N \cdot \ep^N \nonumber \\
\de_{_{\ep}} a_\mu\,^M\,_N &=& \pa_\mu \ep^M\,_N + \ep^M\,_R \cdot a_\mu\,^R\,_N - a_\mu\,^M\,_R \cdot \ep^R\,_N \\
&+& 2\,a_\mu\,^R \cdot \ep^M\,_{RN} - 2\, \ep^R \cdot a_\mu\,^M\,_{RN}
\nonumber \\
&\dots& \nonumber
\eeqq
Note that the last line vanishes if we only keep the $E({\bf R}^D)$-transformations and -fields.

Let us next expand the field strength components
\beqq \label{25}
F_{\mu\nu}\,^M (x,X) &=& f_{\mu\nu}\,^M (x) 
+ f_{\mu\nu}\,^M\,_R (x)\,X^R \nonumber \\
&+& f_{\mu\nu}\,^M\,_{RS} (x)\,X^R X^S + O(X^3).
\eeqq
Again the first two terms are the translation and rotation field strengths $f_{\mu\nu}\,^M$ and $f_{\mu\nu}\,_{asym}^{MN}$ respectively related to the invariance of Isometrodynamics under the local Euclidean group $E({\bf R}^D)$, where
\beq \label{26}
f_{\mu\nu}\,^{MN} (x) = f_{\mu\nu}\,_{asym}^{MN} (x)
+ f_{\mu\nu}\,_{sym}^{MN} (x),
\quad f_{\mu\nu}\,_{sym}^M\,_M (x) = 0
\eeq
is decomposed in its anti-symmetric and symmetric components. From Eqn.(\ref{5}) we next read off the field strength components in terms of the gauge fields 
\beqq \label{27}
f_{\mu\nu}\,^M &=& \pa_\mu a_\nu\,^M - \pa_\nu a_\mu\,^M
+ a_\nu\,^M\,_N \cdot a_\mu\,^N - a_\mu\,^M\,_N \cdot a_\nu\,^N \\
f_{\mu\nu}\,^M\,_N &=& \pa_\mu a_\nu\,^M\,_N - \pa_\nu a_\mu\,^M\,_N
+ a_\nu\,^M\,_R \cdot a_\mu\,^R\,_N
- a_\mu\,^M\,_R \cdot a_\nu\,^R\,_N \nonumber \\
&+& 2\, a_\mu\,^R \cdot a_\nu\,^M\,_{RN} - 2\, a_\nu\,^R \cdot a_\mu\,^M\,_{RN} \nonumber
\eeqq
and from Eqn.(\ref{6}) their variation under a gauge transformation
\beqq \label{28}
\de_{_{\ep}} f_{\mu\nu}\,^M &=& \ep^M\,_N \cdot f_{\mu\nu}\,^N - f_{\mu\nu}\,^M\,_N \cdot \ep^N \nonumber \\
\de_{_{\ep}} f_{\mu\nu}\,^M\,_N &=& \ep^M\,_R \cdot f_{\mu\nu}\,^R\,_N - f_{\mu\nu}\,^M\,_R \cdot \ep^R\,_N \\
&+& 2\, f_{\mu\nu}\,^R \cdot \ep^M\,_{RN}
- 2\, \ep^R \cdot f_{\mu\nu}\,^M\,_{RN} \nonumber \\
&\dots& \nonumber 
\eeqq
Note that each $x$- and $X$-dependent field is equivalent to an infinite string of purely $x$-dependent coefficient fields. Setting all these coefficient fields but the translation and rotation modes to zero reduces Isometrodynamics to the Yang-Mills gauge field theory of $E({\bf R}^D)$.

In order to derive the dynamics of the effective fields arising in the Taylor expansion Eqn.(\ref{22}) of the ID gauge fields we turn to inserting the expansion Eqn.(\ref{25}) of the field strength components into the action of ID and integrating over "inner" space.

Starting with the Lagrangian of ID Eqn.(\ref{19}) in a Euclidean gauge ($g_{MN} = \de_{MN}$)
\beq \label{29}
L_{ID}
= - \int \!d^D X\, \lp^{-D} \frac{1}{4 g^2}\, \frac{c^4}{\Ga} \: F_{\mu\nu}\,^M \cdot F^{\mu\nu}\,_M 
\eeq
and inserting the expansion for the product of the field strength components
\beqq \label{30}
F_{\mu\nu}\,^M \cdot F^{\mu\nu}\,_M
&=& f_{\mu\nu}\,^M \cdot f^{\mu\nu}\,_M 
+ 2\, f_{\mu\nu}\,^M \cdot f^{\mu\nu}\,_{MR}\,X^R \nonumber \\
&+& f_{\mu\nu}\,^M\,_R \cdot f^{\mu\nu}\,_{MS}\,X^R X^S \\
&+& 2\, f_{\mu\nu}\,^M \cdot f^{\mu\nu}\,_{MRS}\,X^R X^S 
+O(X^3) \nonumber
\eeqq
we integrate over "inner" space and finally obtain the effective Lagrangian for the various $x$-dependent field modes 
\beqq \label{31}
L_{ID} &=& - \frac{1}{4 g^2}\, \frac{c^4}{\Ga} \Bigg\{
f_{\mu\nu}\,^M \cdot f^{\mu\nu}\,_M
+ \frac{\lp^2}{12}\, f_{\mu\nu}\,^M\,_R \cdot f^{\mu\nu}\,_M\,^R
\nonumber \\
&+& \frac{\lp^2}{6}\, f_{\mu\nu}\,^M \cdot f^{\mu\nu}\,_M\,^R\,_R
+ O(\lp^4) \Bigg\}.
\eeqq
This effective Lagrangian defines the dynamics of the infinite string of coefficient fields in the effective field expansion.

To get this result we have in effect assumed that all fields are concentrated in
"inner" space in a $D$-dimensional cube of length $\lp$ and vanish outside, i.e. that "inner" space has a finite volume 
\beqq \label{32} 
\int_{\mid X^R\mid \leq \frac{\lp}{2}}
\,d^{D}X &=& \lp^D, \nonumber \\
\int_{\mid X^R\mid \leq \frac{\lp}{2}}
\,d^{D}X \, X^R &=& 0, \\
\int_{\mid X^R\mid \leq \frac{\lp}{2}}
\,d^{D}X \, X^R X^S
&=& \frac{1}{12}\, \lp^{D+2}\, \de^{RS}. \nonumber
\eeqq
This is an important additional assumption which will bear interesting consequences - on the one hand it is the basis of the technical verification of criterium (iii) above, on the other it will point to masses of fundamental particles being quantized, i.e. being multiples of a fundamental unit mass as we will argue below. Accordingly from now own we will work with the gauge group ${\overline{DIFF}}\,[-\frac{\lp}{2},\frac{\lp}{2}]^D$ of diffeomorphisms preserving the $D$-dimensional cube $[-\frac{\lp}{2},\frac{\lp}{2}]^D$.

By construction the dependence of the Lagrangian Eqn.(\ref{31}) on the "inner" space variables has completely dissappeared and the resulting effective field theory is organized in terms of even powers of the Planck length. As $\lp^2 = \frac{\hbar \Ga}{c^3}$ all but the leading term vanish in the limit $\hbar\ar 0$ and are of true quantum nature. In other words the effective field theory reduces in the classical limit to the leading term in the expansion which in turn determines the translation mode dynamics which becomes Abelian in this limit. Note that strictly speaking there is no classical limit of the pure gauge theory because of asymptotic freedom (and its flip side confinement). Only in the presence of the Standard Model fields QID is not asymptotically free for $2\leq D \leq 6$. In the following we will assume the SM fields present and $2\leq D \leq 6$ so that asymptotic ID states and a classical limit exist.

\section{Coupling to Matter and Classical Point Particle Equations}
In this section we analyze a simple model - Isometrodynamics coupled to a Dirac spinor (and with all SM fields in the background to avoid confinement). Expanding the fields as in the previous section and retaining the leading order terms only we show that the translation mode of the effective field theory couples to the conserved Noether current of the Dirac field related to global translation invariance in "inner" space. In complete analogy to Electrodynamics we then write down the relativistic Lagrangian of a classical point particle carrying ID charge $K^M$ interacting with the translation mode $a_\mu\,^M$ of the ID gauge field.

Let us start with the Lagrangian density of a free Dirac spinor in SI units
\beq \label{33}
{\cal L}_D \equiv – {\overline\psi} \, \left[\hbar c\, \ga^\mu \pa_\mu + m c^2\right] \psi.
\eeq
The minimal coupling prescription requires us to re-define the Dirac field $\psi (x)\ar \PSI (x,X)$ on ${\bf M\/}^{\sl 4}\times
[-\frac{\lp}{2},\frac{\lp}{2}]^{D}$ - adding the necessary additional "inner" degrees of freedom - and to replace ordinary derivatives through covariant ones $\pa_\mu\ar D_\mu = \pa_\mu + A_\mu\,^M\cdot \nabla_M$ in the Lagrangian above. As a result we get
\beq \label{34}
{\cal L}_D = – {\overline\PSI}\, \left[\hbar c\, \ga^\mu (\pa_\mu + A_\mu\,^M\cdot \nabla_M) + m c^2\right] \PSI.
\eeq

The coupled system is a simple model with the action
\beqq \label{35}
& & S = - \int \!dt\int \!d^3 x\int \!d^D X\, \lp^{-D}
\Bigg\{ \,\frac{1}{4 g^2}\, \frac{c^4}{\Ga} \: F_{\mu\nu}\,^M
\cdot F^{\mu\nu}\,_M \nonumber \\
& & \quad\quad\quad\quad +\, {\overline\PSI}\, \left[\hbar c\, \ga^\mu (\pa_\mu + A_\mu\,^M\cdot \nabla_M) + m c^2\right] \PSI \Bigg\}.
\eeqq
Note that if the Dirac field would be constant over "inner" space it would automatically decouple from ID because ID fields couple to field variations over "inner" space only. We then could integrate over "inner" space and would recover in effect the action Eqn.(\ref{33}) we started with before minimal coupling. The same holds true for any field theory such as the Standard Model. It is in this sense that in the limit of ID being "turned off" the non-ID laws of physics reduce to the laws of special relativity - technically validating criterium (iii).

Expanding next the ID gauge field as in the previous section, retaining only the leading translation mode and focusing on the interaction term in the action Eqn.(\ref{35}) above we find
\beqq \label{36}
S_{INT} &=& - \int \!dt\int \!d^3 x \, a_\mu\,^M\cdot
\int \!d^D X\, \lp^{-D} \, {\overline\PSI}\,\hbar c\, \ga^\mu \nabla_M \PSI \nonumber \\
&=& - \int \!dt\int \!d^3 x \, a_\mu\,^M\cdot j^\mu\,_M.
\eeqq
Here
\beq \label{37}
j^\mu\,_M = \int \!d^D X\, \lp^{-D} \, {\overline\PSI}\, \hbar c\, \ga^\mu \nabla_M \PSI
\eeq
is the conserved Noether current related to the global translation invariance of the Lagrangian density Eqn.(\ref{34}) as discussed in detail in \cite{chw2}. Note that this current coupling to the gauge field carries the dimension of an energy density, yet is not the usual energy-momentum tensor.

In Electrodynamics to get the classical description of a point charge corresponding to the quantum description based e.g. on Dirac's equation coupled to the electromagnetic field the Noether current $j^\mu = e\, {\overline\psi}\, \hbar c\, \ga^\mu \psi $ of the quantum theory related to global $U(1)$-invariance is replaced by a point charge $e$ carried along the classical trajectory $y^\mu (\tau)$ of the point particle $j^\mu (x)\ar c\, e \int d\tau\, \dot y^\mu\, \de^4 (x-y(\tau))$ with $\tau$ denoting the proper time.

In tentative analogy we replace Eqn.(\ref{37}) by
\beq \label{38}
j^\mu\,_M (x)\ar c\, K_M \int d\tau\, \dot y^\mu\, \de^4 (x-y(\tau)),
\eeq
in effect turning to a description of a classical point particle carrying the conserved "inner" momentum $c\, K_M = \int d^3 x\, j^{\sl 0}\,_M (t,{\bf x})$ along its trajectory. This is a good approximation at least for currents $j^\mu\,_M (x)$ strongly concentrated in three-dimensional space around $y(\tau)$. However, this approximation destroys the local nature of the Noether current - which will come at an important price as we will see below.

The effective action for the theory coupling the translation field mode and the classical point particle with rest mass $m_{_K}$ becomes
\beqq \label{39}
S &=& - \int \!dt\int \!d^3x \,\frac{1}{4 g^2}\, \frac{c^4}{\Ga} \: f_{\mu\nu}\,^M\cdot f^{\mu\nu}\,_M \nonumber \\
&-& \int d\tau\, \frac{m_{_K}}{2}\,\dot y_\mu\, \dot y^\mu 
- \int d\tau\, a_\mu\,^M\cdot K_M\, \dot y^\mu 
\eeqq
which is the core result of this section.

Let us finally check the gauge-invariance of Eqn.(\ref{39}) under the leading term of Eqn.(\ref{20}). The first two terms are obviously gauge-invariant whereas for the third we find 
\beqq \label{40}
\de_{_{\cal E}} S_{INT}
&=& -\int d\tau\, \pa_\mu\ep^M\cdot K_M\, \dot y^\mu \nonumber \\
&=& -\int d\tau\, \frac{d \ep^M}{d \tau}\cdot K_M = 0 
\eeqq 
if the gauge parameter $\ep^M (y(\tau_1)) = \ep^M (y(\tau_2)) = 0$ vanishes at the boundaries. So Eqn.(\ref{39}) defines an Abelian gauge theory, though not based on the usual U(1) Abelian gauge group.

\section{Leading Order Solution of the Coupled System}
In this section we solve the coupled system of a classical point particle interacting with the translation mode of the effective field theory and obtain potentials analogous to the Lienard-Wiechert potentials in Electrodynamics.

From Eqn.(\ref{39}) it is easy to derive the equations of motion for the combined system
\beqq \label{41}
\frac{c^4}{\Ga} \: \pa_\mu f^{\mu\nu}\,_M
&=& g^2\, c\, K_M \int d\tau\, \dot y^\nu\, \de^4 (x-y(\tau)) \\
\label{42} m_{_K}\, \ddot y_\mu &=& f_{\mu\nu}\,^M\cdot K_M\, \dot y^\nu.
\eeqq
Note that in this approximation all the higher modes $a_\mu\,^M\,_R$, $a_\mu\,^M\,_{RS}$, $\dots$ vanish and the effective theory linearizes.

In the Lorentz gauge $\pa_\mu a^\mu\,_M = 0$ Eqn.(\ref{41}) becomes
\beq \label{43}
\frac{c^3}{\Ga}\, \pa^2 a^\nu\,_M
= g^2\, K_M \int d\tau\, \dot y^\nu\, \de^4 (x-y(\tau))
\eeq
which is easily solved using the retarded Green function
\beq \label{44}
G_{ret} (x) = -\frac{1}{2\pi}\, \theta(x_{\sl 0})\, \de(x^2)
\eeq
fulfilling $\pa^2 G_{ret} (x-y) = \de^4(x-y)$. As a result we obtain the Lienard-Wiechert potentials for Isometrodynamics
\beqq \label{45}
a_{ret}^\mu\,_M (x)
&=& -\frac{g^2}{2\pi}\, \frac{\Ga}{c^3}\, K_M
\int d\tau\, \theta(x^{\sl 0} - y^{\sl 0}(\tau))\,
\de\left((x - y(\tau))^2\right)\, \dot y^\mu (\tau) \nonumber \\
&=& -\frac{g^2}{4\pi}\, \frac{\Ga}{c^3}\, K_M\,
\frac{\dot y^\mu_+}{\dot y_+ \cdot(x - y_+)}.
\eeqq
Above we have introduced $y_+ = y(\tau_+)$ which is the uniquely determined retarded point on the particle trajectory with
\beq \label{46}
(x - y_+)^2 = 0,\quad y_+^{\sl 0} < x^{\sl 0}.
\eeq 
For ${\bf r} = {\bf x} - {\bf y}_+$ and $\hat{\bf v} =
\frac{1}{c}\, \frac{d{\bf y}(\tau_+)}{dt}$ the potentials read more explicitly 
\beqq \label{47}
a^{\sl 0}\,_M (x) &=& \frac{g^2}{4\pi}\, \frac{\Ga}{c^3}\, K_M\,
\frac{1}{r - {\bf r}\cdot \hat{\bf v}} \\
\label{48} {\bf a}_M (x) &=& \frac{g^2}{4\pi}\, \frac{\Ga}{c^3}\, K_M\,
\frac{\hat{\bf v}}{r - {\bf r}\cdot \hat{\bf v}},
\eeqq
where bold letters indicate three-vectors.

In analogy to Electrodynamics we next introduce the Isometro-electric and Isometro-magnetic field components $E^i\,_M$ and $B^i\,_M$ respectively 
\beq \label{49}
E^i\,_M \equiv F^{i{\sl 0}}\,_M, \quad
B^i\,_M \equiv -\frac{1}{2}\,\ep^{ijk} F_{jk\,M},
\eeq
where $\ep^{ijk}$ is the totally antisymmetric tensor in three dimensions.

The effective Isometro-electric and -magnetic field components to leading order are then easily found to be
\beqq \label{50}
& & {\bf e}_M = -\,\frac{g^2}{4\pi}\, \frac{\Ga}{c^3}\, K_M\,
\frac{({\bf r} - r \hat{\bf v})(1 - {\hat v}^2) + {\bf r}\wedge \left(({\bf r} - r\hat{\bf v})\wedge \frac{1}{c}\, \frac{d\hat{\bf v}}{dt}\right)}
{( r - {\bf r}\cdot \hat{\bf v})^3} \quad\quad\quad\quad \\
& & \label{51} \quad\quad\quad\quad\quad\quad\quad\quad\quad\quad
{\bf b}_M = \frac{\bf r}{r} \wedge {\bf e}_M 
\eeqq
and reduce to
\beq \label{52}
{\bf e}_M = -\,\frac{g^2}{4\pi}\, \frac{\Ga}{c^3}\, K_M\,
\frac{{\bf r}}{r^3},\quad {\bf b}_M = 0
\eeq
in the static approximation $\hat{\bf v} = 0$. The field components are in complete analogy to Electrodynamics and there is radiation by accelerated point particles in the effective field theory - terms $\propto \frac{1}{r^2}$ only contribute at short distance and the long-ranging terms $\propto \frac{1}{r}$ in Eqn.(\ref{50}) are the ones related to radiation. The formal analogy of the effective field approximation to leading order between Isometro- and Electrodynamics can be used to easily transfer powerful techniques such as the multipole expansion to the effective field theory.

\section{Newton's Inverse Square Law}
In this section identifying the lengths of the "inner" momentum vectors with the point particle rest masses and taking the two "inner" momentum vectors of the point particles generating and sensing the translation field mode respectively as anti-parallel to fix one of the parameters of the effective field theory we obtain Newton's inverse square law.

To make a further step in tentatively interpreting ID let us turn to the motion of a point particle $A$ with trajectory $z^\mu$ and "inner" momentum $Q^M$ interacting with the fields Eqns.(\ref{50}) and (\ref{51}) generated by another point particle $B$ with trajectory $y^\mu$ and "inner" momentum $K^M$ within the realm of the effective field theory developed in the previous sections. We are specifically interested in the limiting case where $B$ is static and $A$ moves slowly and find from Eqns.(\ref{42}) and (\ref{52}) with $\dot z^\mu \approx (c, \frac{d {\bf z}}{dt})$ for this case
\beq \label{53}
m_{_K}\, \frac{d^2 {\bf z}}{dt^2} = \frac{g^2}{4\pi}\, \frac{\Ga\, K_M\, Q^M}{c^2}\,
\frac{{\bf z}}{\mid\! {\bf z}\!\mid^3}.
\eeq

Note that both $K$ and $Q$ are Euclidean vectors in ${\bf R\/}^{D}$ so that their scalar product can be expressed as
\beq \label{54}
K_M\, Q^M = \mid\! K\!\mid \mid\! Q\!\mid \cos\theta,
\eeq
where $\theta$ is the angle between the two vectors and $\mid\! K\!\mid\, > 0$, $\mid\! Q\!\mid\, > 0$ carry the dimension of momentum.

If we tentatively identify $\mid\! K\!\mid\, = m_{_K} c$ and $\mid\! Q\!\mid\, = m_{_Q} c$, where $m_{_K}$ is the inertial mass of the point particle $B$ and $m_{_Q}$ the inertial mass of the point particle $A$, we find
\beq \label{55}
m_{_K}\, \frac{d^2 {\bf z}}{dt^2} = \frac{g^2}{4\pi}\, \Ga\, m_{_K}\, m_{_Q}\, \cos\theta\, \frac{{\bf z}}{\mid\! {\bf z}\!\mid^3}.
\eeq 

Apart from the factor $\frac{g^2}{4\pi}$ which can be absorbed in a rescaled $g$ we find a force law which depends on the relative direction of $K$ and $Q$ and which reduces exactly to Newton's law if $K$ is anti-parallel to $Q$, i.e. choosing the effective field theory parameter $\cos\theta = -1$.

Hence, the effective field theory developed in the previous sections approximating Isometrodynamics can reproduce Newton's law in a natural way if we interpret the "inner" vectors $K$, $Q$, $\dots$ as gravitational mass vectors with lengths identical to the inertial masses of the respective point particles $m_{_K} c$, $m_{_Q} c$, $\dots$. Hence an interpretation of ID in terms of gravitational physics leads to the notion of gravitational mass having vector character in "inner" space.

Ultimately the directions of these "inner" mass vectors $K$, $Q$ should be determined by the full local dynamics of ID - in the approximation analyzed here they are fixed and determine the force between two point particles to be either attractive or repulsive with Newton's law resulting in the limiting case of $K$ being anti-parallel to $Q$. This is the price we pay for replacing the local Noether currents coupling to the gauge fields by non-local charges carried along point-particle trajectories.

As noted it is the full dynamics which should determine the relative directions of $K$ and $Q$ - but at least a heuristic argument for $\cos\theta = -1$ can be established looking at the interaction energy in the static approximation
\beq \label{56}
{\cal H}_{INT} = \frac{g^2}{4\pi}\, \frac{\Ga\, K_M\, Q^M}{c^2}\, \frac{1}{r}
= \frac{g^2}{4\pi}\, \frac{\Ga\, m_{_K}\, m_{_Q} }{r}\, \cos\theta.
\eeq
Viewed as a function of $\cos\theta$ the energy is minimized for $\cos\theta = -1$ or $K$ being anti-parallel to $Q$, hence in the Newton case.

As a non-trivial additional result the effective theory automatically yields a universal trajectory for all classical point particles moving in the effective translation mode field - technically Eqn.(\ref{42}) can be rewritten as
\beq \label{57}
\ddot z_\mu = f_{\mu\nu}\,^M\, \frac{Q_M}{m_{_Q}}\, \dot z^\nu
\eeq
which is independent of $m_{_Q}$. We have recovered the Weak Equivalence Principle in the limit of classical point particles.

Core to get the correct Newtonian limit have been the identification of the length of the conserved "inner" momentum vector with the particle mass - in effect representing gravitational mass in Isometrodynamics as a Euclidean vector - and the possibility to set the angle between two such vectors, the effective parameter $\cos\theta = -1$.

Note that the identification of "inner" momentum vectors with the particle mass might indicate mass quantization - for the simple fact that if "inner" space is compact the "inner" momentum vectors being Fourier conjugates of the "inner" space points $X$ become discrete. Indeed for a generic matter quantum $a^\dagger (k,K; m_{_K}) \mid\! 0\rangle$ generated from the vaccuum we can measure the "inner" momentum of the state vector
\beqq \label{58}
{\bf K}^M\, a^\dagger (k,K; m_{_K}) \mid\! 0\rangle &=& [{\bf K}^M\, ,a^\dagger (k,K; m_{_K})] \mid\! 0\rangle \nonumber \\
&=& K^M\, a^\dagger (k,K; m_{_K}) \mid\! 0\rangle,
\eeqq
where ${\bf K}^M$ denotes the "inner" momentum operator. Because of the compactness of "inner" space we find for the square of the "inner" momentum
\beq \label{59}
K_M\, K^M = m_{_K}^2 c^2\propto \frac{\hbar^2\pi^2}{l_P^2}\cdot\!\!\!\sum_{M=1,\dots,D} n_{M{_K}}^2,\quad\quad n_{M{_K}}\in {\bf N\/}^{D}
\eeq
after identification of $\mid\! K\!\mid\, = m_{_K} c$. As a result the masses $m_{_K}$ squared of one-particle states corresponding to fundamental fields might turn out to be integer multiples of the Planck mass $\mp$ squared $m_{_K}^2\propto \mp^2 \cdot\sum_{M=1,\dots,D} n_{M{_K}}^2$. This indeed points to mass quantization.

There is one important open point related to the identification of $\mid\! K\!\mid\, = m_{_K} c$. For general state vectors $a^\dagger (k,K; m_{_K}) \mid\! 0\rangle$ spanning the one-particle asymptotic Hilbert space $k$ and $K$ will not be related as is the case for the states with $\mid\! K\!\mid\, = m_{_K} c$. So the incoming and outgoing asymptotic one-particle states which can be identified with observable point particles on the basis of $\mid\! K\!\mid\, = m_{_K} c$ form only a subset of the full asymptotic Hilbert space. As a consequence the relation of general one-particle QFT states and observable point particles has to be further clarified.

Note finally that we have two different conservation laws for theories defined on ${\bf M\/}^{\sl 4}\times {\bf R\/}^{D}$ - energy momentum conservation related to spacetime translation invariance and "inner" momentum conservation related to global "inner" translation invariance. And both are related to the invariant particle masses. Could this lead to additional, experimentally verifiable kinematical restrictions on processes such as two-body scattering?

\section{Effective Field Theory and Spacetime Geometry}
In this section we analyze the behaviour of rods and clocks in the presence of effective translation field modes and discuss the connection of ID with the General Theory of Relativity (GR).

To do so let us analyze the line element measuring invariant spacetime distances at the spacetime point $y$ in the presence of gauge fields. Note that "invariant" in the context of ID means gauge-invariant and not invariant under general coordinate transformations.

The line element has to be the gauge-invariant generalization of the line element at point ${\bf y}$ in the absence of gauge fields
\beq \label{60}
ds^2 = -\, dy_\mu\, dy^\mu = -\, {\dot y}_\mu\, {\dot y}^\mu\, d\si^2,
\eeq 
where $\si$ is an invariant parameter along the spacetime trajectory to be measured and $\dot {}$ denotes a derivative w.r.t. $\si$.

Noting that it is always possible to gauge away ID gauge fields in one point \cite{chw2} the natural gauge-invariant generalization of the line element Eqn.(\ref{60}) in the presence of an effective translation background field is
\beq \label{61}
ds^2 = -\, \left(\dot y_\mu\, \dot y^\mu + 2\, a_\mu\,^M\cdot \frac{K_M}{m_{_K}}\, \dot y^\mu \right) d\si^2,
\eeq
where the gauge invariance has been shown in Section 6, Eqn.(\ref{40}). This expression does not depend on the specific test body following the trajectory $y^\mu (\si)$ and applies in particular to any rod or clock on the trajectory - the reason it is indeed measuring invariant spacetime distances.
 
For the gauge field Eqn.(\ref{47}) generated by a static point particle with mass $m_{_Q}$ located in the origin of the coordinate system the invariant distance becomes
\beqq \label{62}
ds^2 &=& c^2 \left(1 + \frac{g^2}{2\pi}\, \frac{\Ga\, m_{_Q}}{c^2\, r}\, 
\frac{Q^M K_M}{m_{_Q} c \cdot m_{_K} c} \right) dt^2  -\, d{\bf y}^2 \nonumber \\
&=& c^2 \left(1 -\, 2\, \frac{\Ga\, m_{_Q}}{c^2\, r} \right) dt^2  -\, d{\bf y}^2,
\eeqq
where we have made use of $Q^M K_M = m_{_Q} c \cdot m_{_K} c\, \cos\theta$ and took the parameters $\cos\theta = -1$ and $\frac{g^2}{4\pi} = 1$ in the effective field theory in accordance with the Newtonian limit.

This means that clocks in the presence of a static Newtonian potential Eqn.(\ref{47}) tick slower or more generally that the spacetime metric measured in experiments is not flat - reproducing a core result of General Relativity \cite{stw4,ros}. Note that this fact does not destroy the fundamental role of the Minkowski metric played in the very development of Isometrodynamics as a field theory formulated on flat spacetime \cite{chw2,chw3} - essentially because the fundamental fields are the ID gauge fields and the measured spacetime metric is just one of many gauge-invariant observables derived from the fundamental fields.

\section{Radiation of Classical Point Particles in the Effective Field Theory}
In this section we determine the gravitational radiation of a classical point particle in the effective field theory approximation and calculate as an illustration the power radiated by earth on its orbit around the sun.

Let us start with the gauge-invariant and symmetric energy-momentum tensor for Isometrodynamics \cite{chw2} in SI units
\beq \label{63}
\!\!\!\!\Th^\mu\,_\nu = -\!\int \!d^D X\, \lp^{-D}
\frac{1}{4 g^2} \frac{c^4}{\Ga}
\left\{\frac{1}{4}\,\eta^\mu\,_\nu\, F_{\rho\si}\,^M\! \cdot F^{\rho\si}\,_M - \,F^{\mu\rho}\,_M\! \cdot F_{\rho\nu}\,^M \right\}.
\eeq
As shown in \cite{chw2} this tensor is conserved
\beq \label{64}
\pa_\mu\Th^\mu\,_\nu = 0.
\eeq
The corresponding time-independent momentum four-vector
\beq \label{65} P_\mu = \int\! d^{\sl 3}x\, \Th^{\sl 0}\,_\mu \eeq
fulfills
\beq \label{66}
-\, \pa_{\sl 0} P_\mu  =  \int\! d^{\sl 3}x\, \pa_i \Th^i\,_\mu.
\eeq
We now rewrite Eqn.(\ref{63}) for $\nu = 0$ with the use of the Isometro-electric and Isometro-magnetic field components $E^i_M$ and $B^i_M$ Eqns.(\ref{49})
\beqq \label{67}
& & \Th^{\sl 0}\,_{\sl 0} = \int \!d^D X\, \lp^{-D}
\frac{1}{8 g^2} \frac{c^4}{\Ga}
\left( {\bf E}^M\cdot {\bf E}_M + {\bf B}^M\cdot {\bf B}_M \right) > 0, \nonumber \\
& & \quad\quad\quad\quad \Th^i\,_{\sl 0} = \int \!d^D X\, \lp^{-D}
\frac{1}{4 g^2} \frac{c^4}{\Ga}
\left( {\bf E}^M \wedge {\bf B}_M \right)^i,
\eeqq
where we have explicitly calculated both the energy density $\Th^{\sl 0}\,_{\sl 0}$ to check the correct normalization and the Poynting vector $\Th^i\,_{\sl 0}$ in Isometrodynamics. Insertion in Eqn.(\ref{66}) yields
\beq \label{68}
-\, \frac{1}{c} \frac{dP_{\sl 0}}{dt} = \int\! d^{\sl 3}x\, \pa_i\! \int \!d^D X\, \lp^{-D} \frac{1}{4 g^2} \frac{c^4}{\Ga}
\left( {\bf E}^M \wedge {\bf B}_M \right)^i.
\eeq

In the following we work with the leading order Isometro-electric and Isometro-magnetic translation field mode which depends on spacetime coordinates only and is constant over "inner" space.

Integrating over "inner" space and using Gauss's theorem we find
\beq \label{69}
-\, \frac{1}{c} \frac{d P_{\sl 0}}{dt} 
= \frac{c^4}{\Ga}\, \int\! d {\bf S}_i 
\left( {\bf e}^M \wedge {\bf b}_M \right)^i,
\eeq
where $d {\bf S}\equiv {\bf n}\,\, r^2\, d\Om$ with ${\bf n} \equiv \frac{\bf r}{r}$ is the directed two-dimensional surface element.

Taking the explicit form of the Isometro-electric and Isometro-magnetic field components Eqns.(\ref{50}) and (\ref{51}) in the wave zone $\propto \frac{1}{r}$
\beq \label{70}
{\bf e}^M = \frac{\Ga}{c^4}\,\, K^M\, 
\frac{{\bf n}\wedge \left(({\bf n} - \hat{\bf v})\wedge
\frac{d\hat{\bf v}}{dt}\right)}{( 1 - {\bf n}\cdot \hat{\bf v})^3}\,\, \frac{1}{r},
\eeq 
where we have set $\frac{g^2}{4\pi} = 1$ to be in accordance with the correct Newtonian limit as discussed in the previous section, we find
\beq \label{71}
{\bf e}^M \wedge {\bf b}_M = \left({\bf e}^M\cdot {\bf e}_M\right) {\bf n}.
\eeq

Defining $P_{ret} \equiv \frac{1}{c} \frac{d P_{\sl 0}}{\,\,\,\,dt_{ret}}$ the radiated angular power distribution is then easily found to be
\beq \label{72}
-\, \frac{d P_{ret}}{d\Om}
= \frac{\Ga}{c^3}\, m_{_K}^2 c^2\,
\frac{\mid\! {\bf n}\wedge \left(({\bf n} - \hat{\bf v})\wedge
\frac{d \hat{\bf v}}{\,\,\,\,dt_{ret}}\right)\!\!\mid ^2}{( 1 - {\bf n}\cdot \hat{\bf v})^5},
\eeq
where $m_{_K}$ denotes the mass of the radiating point particle. Note that a factor of $\frac{dt}{\,\,\,\,dt_{ret}}= 1 - {\bf n}\cdot \hat{\bf v}$ has decreased the exponent in the denominator by one. Integrating out the angles leaves us with the relativistic Larmor formula for Isometrodynamics 
\beq \label{73}
- P_{ret} = \frac{8\pi}{3}\, \frac{\Ga}{c^3}\, m_{_K}^2 c^2\,
\frac{ \mid\!\! \frac{d \hat{\bf v}}{\,\,\,\,dt_{ret}}\!\!\mid ^2 -
\mid\! \hat{\bf v} \wedge \frac{d \hat{\bf v}}{\,\,\,\,dt_{ret}}
\!\!\mid ^2}{( 1 - \hat{\bf v}^2)^3},
\eeq

Note that accelerated point particles in Isometrodynamics come along with dipole radiation as they do in Electrodynamics - a prediction which distinguishes ID clearly from GR \cite{cmw,stw4,ros}.

As an application let us determine the power radiated by the earth's accelerated orbit around the sun. Approximating the trajectory by a cercle with radius $\rho$ the formula Eqn.(\ref{73}) further simplifies to
\beq \label{74}
- P_{ret} = \frac{8\pi}{3}\,\Ga\, m_{_K}^2\,\frac{c}{\rho^2}\,
\frac{\hat{\bf v}^4}{( 1 - \hat{\bf v}^2)^2}.
\eeq

Putting numbers into Eqn.(\ref{74}) results in a power of 26.4 Gigawatts radiated off by the earth.

\section{Conclusions}

In this paper we have proposed a tentative interpretation of Isometrodynamics in terms of gravitational physics on the basis of ID fullfilling central requirements towards any theory of gravity. In particular ID fullfils universality - i.e. the automatic coupling of gravity to all fundamental fields - which we believe is essential to the nature of gravity.

In GR universality is built into the theory in terms of spacetime geometry, i.e. the universal trajectories of classical particles are taken as geodesics and can be interpreted as directly related to spacetime and its geometry irrespective of the inner composition of a test particle. The task in GR is then to find the relation between spacetime geometry and matter, which results in Einstein's equations. A heuristic generalization invoking the principle of equivalence finally allows to write down the coupling of non-gravitational fields to the spacetime metric \cite{cmw,stw4,ros}.

In ID universality is built into the theory in gauge-symmetry terms, i.e by gauging the group of volume-preserving diffeomorphisms represented on a $D$-dimensional "inner" space attached to each spacetime point. This results automatically in an universal coupling of the ID gauge fields to any non-gravitational field. The task is now to define the dynamics of the ID field which has been tackled in \cite{chw2}. In this emerging picture gravity couples to {\it variations} of the non-gravitational fields over "inner" space. Turning off gravity then amounts to only keeping the mode of any non-gravitational field constant over "inner" space.

To technically develop the program above our starting point has been the dimensional analysis of pure Isometrodynamics which introduces the Planck mass and length in a natural way into the theory and which fixes the unique combination in which $c$ and notably $\Ga$, but not $\hbar$, enter the Lagrangian density of ID.

Noting that ID couples solely through derivatives acting on "inner" space variables - hence only to fields varying, but not to fields constant over "inner" space - we have Taylor-expanded all the gauge parameters, gauge fields and field strengths of ID in the "inner" space variable $X$. This expansion in terms of $X$-derivatives amounts to a weak coupling approximation. Note that the resulting infinite string of spacetime dependent coefficient fields in the expansion corresponds one-to-one to the full ID fields depending on both spacetime and "inner" space variables.

Restricting the full fields to live inside a finite-volume cube of tentative length $l_P$ of "inner" space, truncating the expansion at $O(X^2)$ and integrating out the "inner" space variables has resulted in an effective field theory for the lowest expansion modes organized in even powers of the Planck length - or $\hbar$ - with the leading mode corresponding to the "inner" global translation symmetry of the full theory. In the limit $\hbar \ar 0$ only the dynamics for this leading mode does not vanish - singleing out the translation mode and its dynamics as the classical remnant of full ID. 

Turning to a model with ID coupled to a Dirac spinor we have found that the translation mode of the ID gauge field expansion couples to the conserved Noether current related to the "inner" global translation invariance of the Dirac Lagrangian. In analogy to Electrodynamics we then have approximated this current by a point particle carrying along the corresponding Noether charge, in the ID case the conserved "inner" momentum $K$. As a result we have obtained an effective theory for the translation mode coupled to a classical point particle carrying "charge" $K$. This approximation should make sense for slowly varying ID fields interacting with "matter" field currents strongly concentrated around a point particle trajectory - hence, in a classical world. The resulting effective theory is solvable for the gauge fields generated by relativistic point sources in complete analogy to Electrodynamics yielding the Lienard-Wiechert potentials for ID.

In an additional approximation we then have looked at an ID field generated by a static point source and a test particle slowly moving in that field. The resulting equation of motion suspiciously looks like Newton's equation of motion for a particle moving in the Newtonian gravitational field generated by a static point mass. Indeed, interpreting the "inner" vectors $K$, $Q$ of the two particles involved as "gravitational mass vectors" with lengths identical to their inertial masses $m_{_K} c$, $m_{_Q} c$ - and taking them to be anti-parallel in "inner" space - yields Newton's inverse square law for the gravitational force between two point masses.

There are a few potentially far reaching implications from the identification of gravitational mass with the "inner" vectors $K$ the most important of which is the possibility of mass quantization related to the discrete nature of the "inner" vectors $K$ following from the assumed compactness of "inner" space. Second there are the implications following from the two different conservation laws for energy-momentum and for the "inner" vectors $K$ related to the inertial mass. And a third one is the universality of the trajectories of point masses moving in translation mode fields – which is equivalent to the validity of the weak principle of equivalence in the effective field theory approximation to leading order.

Finally we have analyzed the potential consequences of our interpretation of ID as a theory of gravity for the spacetime geometry and for point particle radiation. We have found that spacetime geometry looks curved to a classical observer the same way as GR predicts - at least for weak Newtonian-type fields. In contrast to GR though we have found that theres is dipole ID radiation of accelerated point sources - GR would predict the occurrence of quadrupole radiation only.

In summary the following tentative picture is emerging: what we experience macroscopically as gravity is the translation mode of the ID gauge field which couples to matter through the Noether current related to "inner" translation invariance or - for matter being strongly concentrated in small spacetime volumes - through the corresponding Noether charge which is the  "inner" momentum vector $K$ playing the role of "gravitational mass". This reproduces Newton's inverse square law for the gravitational force between two point masses. Macroscopically all other degrees of freedom of the ID gauge field are "frozen" down to reasonably small distances or high momenta of the ID field quanta. Microscopically all modes of the ID field start contributing and recombine to the full ID gauge field obeying a dynamics which by construction looks renormalizable as has explicitly been shown at the one loop level in \cite{chw3}.

Obviously there are many tentative/speculative/unsolved elements in our approach which require further clarification.

On the macroscopic/classical level the full field dynamics should determine that the directions of $K$, $Q$ are anti-parallel in order to yield Newton's law. Then, comparison of the energy loss due to dipole radiation of accelerated particles in ID with the orbital phase shift of the binary pulsar PSR 1913 + 16 should at least be shown not to rule out ID as a candidate theory for gravity. And refining the approximation scheme beyond the linear terms should eventually allow for a derivation of important results such as the perihelion precession of Mercury. 

On the microscopic/quantum level a full prove of renormalizability and the existence of asymptotic states in view of the asymptotic freedom of pure ID have to be established and the values of both the appropriate dimension $D$ of "inner" space and of the coupling constant $g$ have to be determined through anomalous scaling effects - where full Quantum Isometrodynamics will meet experiment. Finally the relation of state vectors in the asymptotic Hilbert spaces with observable particles and classical fields has to be further clarified.

What makes Isometrodynamics an attractive candidate for a consistent classical and quantum theory of gravity in the first place is its structural analogy with the existing gauge field theories of the electromagnetic, weak and strong interactions. If it was the "right" theory we would finally have a unified view of Nature and a consistent framework to describe all fundamental interactions at all scales and without any logical or mathematical rift between the worlds of classical and quantum physics.

\appendix

\section{Notations and Conventions}

Generally, small letters denote spacetime coordinates and parameters, capital letters coordinates and parameters in "inner" space.

Specifically, ({\bf M\/}$^{\sl 4}$,\,$\eta$) denotes $\sl{4}$-dimensional Minkowski spacetime with the Cartesian coordinates $x^\la,y^\mu,z^\nu,\dots\,$ and the spacetime metric $\eta=\mbox{diag}(-1,1,1,1)$. The small Greek indices $\la,\mu,\nu,\dots$ from the middle of the Greek alphabet run over $\sl{0,1,2,3}$. They are raised and lowered with $\eta$, i.e. $x_\mu=\eta_{\mu\nu}\, x^\nu$ etc. and transform covariantly w.r.t. the Lorentz group $SO(\sl{1,3})$. Partial differentiation w.r.t to $x^\mu$ is denoted by $\pa_\mu \equiv \frac{\pa\,\,\,}{\pa x^\mu}$. 
Small Latin indices $i,j,k,\dots$ generally run over the three spatial coordinates $\sl{1,2,3}$ \cite{stw1}.

({\bf R\/}$^{D}$,\,$g$) denotes a $D$-dimensional real vector space with coordinates $X^L, Y^M, Z^N,\dots\,$ and the flat metric $g_{MN}$ with signature $D$. The metric transforms as a contravariant tensor of Rank 2 w.r.t. ${\overline{DIFF}}\,{\bf R}^D$. Because Riem$(g) = 0$ we can always choose global Cartesian coordinates and the Euclidean metric $\de=\mbox{diag}(1,1,\dots,1)$. This is a partial choice of gauge called Euclidean gauge above. The capital Latin indices $L,M,N,\dots$ from the middle of the Latin alphabet run over $\sl{1,2},\dots,D$. They are raised and lowered with $g$, i.e. $X_M=g_{MN} X^N$ etc. and transform as vector indices w.r.t. ${\overline{DIFF}}\,{\bf R}^D$. Partial differentiation w.r.t to $X^M$ is denoted by $\nabla_M \equiv \frac{\pa\quad\,}{\pa X^M}$. 

The same lower and upper indices are summed unless indicated otherwise.

\section*{Acknowledgments}

This paper is dedicated to my parents who have strongly supported my quest for a better understanding of the physics of gravity through my university years and through my time at the Dublin Institute of Advanced Studies.

\end{document}